\documentclass[journal,10pt]{IEEEtran}

\usepackage{graphicx,amssymb,amsmath}
\usepackage{amssymb}
\usepackage{cite}
\usepackage{setspace}
\usepackage{comment}
\usepackage{subfig}

\title{A Scheme of Channel Prediction Based on Artificial Neural Network}
\author{Zirui~Wen, Ruisi~He, Bo~Ai, Chen~Huang, Mi~Yang, and Zhangdui~Zhong

\thanks{Z. Wen is with the China Mobile Research Institute, Beijing, 100053, China (email: wenzirui@chinamobile.com.cn).Part of this work is present in Mr. Zirui Wen's thesis.}
\thanks{R. He, B. Ai, M. Yang, and Z. Zhong are with the State Key Laboratory of Rail Traffic Control and Safety, Beijing Jiaotong University, Beijing 100044, China (email: \{ruisi.he, boai, yangmi\_chn, zhdzhong\}@bjtu.edu.cn).}
\thanks{\noindent C. Huang is with the Pervasive Communication Research Center, Purple Mountain Laboratories, Nanjing, 211111, China; and also with the National Mobile Communications Research Laboratory, School of Information Science and Engineering, Southeast University, Nanjing, 210096, China (email: huangchen@pmlabs.com.cn).}

}
\begin{document}

\maketitle

%
%

\begin{abstract}
  Accurate channel modeling is the foundation of communication system design. However, the traditional measurement-based modeling approach has increasing challenges for the scenarios with insufficient measurement data. To obtain enough data for channel modeling, the Artificial Neural Network (ANN) is used in this paper to predict channel data. The high mobility railway channel is considered, which is a typical scenario where it is challenging to obtain enough data for modeling within a short sampling interval. Three types of ANNs, the Back Propagation Network, Radial Basis Function Neural Network and Extreme Learning Machine, are considered to predict channel path loss and shadow fading. The Root-Mean-Square error is used to evaluate prediction accuracy. The factors that may influence prediction accuracy are compared and discussed, including the type of network, number of neurons and proportion of training data. It is found that a larger number of neurons can significantly reduce prediction error, whereas the influence of proportion of training data is relatively small. The results can be used to improve modeling accuracy of path loss and shadow fading when measurement data is reduced.
\end{abstract}

\section{Introduction}

Channel modeling plays an important role in communication system design. For the traditional channel modeling process, channel measurement is an important approach to obtain data for channel characterization. However, the measurement-based channel modeling is facing various problems, especially in the complex scenarios where channel measurements are challenging. Moreover, with the development of 5G and 6G communications \cite{C}, large bandwidth and antenna array are used, and high mobility is supported. This further makes it challenging to obtain enough data for channel characterization and modeling. Take mobile channel measurement as an example, based on the Lee's theorem\cite{Lee}, to accurately extract the large-scale channel information, at least 36-50 data points need to be obtained in a range of 40 wavelength. This means that if the measurement is conducted in a high mobility scenario and the sampling rate is limited by hardware, it may be challenging to obtain enough data within the 40 wavelength window, and the accuracy of channel characterization (such as path loss and shadow fading modeling) is reduced. Such problem has existed for a long time and the reduced measurement data significantly limits the development of accurate channel modeling. A possible solution is to reconstruct the reduced channel data set using some well-designed algorithms, which can avoid the high cost of repeating channel measurements.\\

Recently, Artificial Intelligence (AI) has been widely considered in the investigation of channel characterization and modeling. The machine learning algorithms can well solve the problems such as clustering, classification, and regression in wireless channel data\cite{w}. In \cite{w44}, the authors used transmitter (Tx) and receiver (Rx) height, distance from Tx to Rx and carrier frequency as Radial Basis Function Neural Network (RBF-NN) input to predict path loss. Ref.\cite{w47} applies an network for modeling and storing ray launching results, and the proposed method achieves high gain in terms of computational efficiency. Ref.\cite{w48} uses Artificial Neural Network (ANN) to model wireless channel and proposes a cluster-nuclei based channel model, and with the measurement data, the prediction of channel is achieved. In \cite{w51}, the received signal strength is predicted by Multi-Layer Perceptron (MLP). The distance from Tx to Rx are regarded as input, and the output is the received signal strength. Ref.\cite{w53} discusses an optimal AI-based approach to fit channel sounder data to a time-variant tapped delay line model. Ref.\cite{w54} proposes an AI-based method to predict coverage and field strength map in the environments where measurements are unavailable. In \cite{w45}, the Relevence Vector Machine (RVM) is used to estimate channel direction of arrival, which obtains signal locations by the sparsity-including RVM on a predefined spatial grid and then obtains refined direction estimation by searching. Moreover, the K-means, Fuzzy C-Means (FCM) and Density-Based Spatial Clustering of Applications with Noise (DBSCAN) have been widely used for channel data clustering \cite{w55}. However, the above existing works mainly use AI to improve channel characterization and modeling, and they actually rely on having a large number of channel data. There is still a lack of effective method to solve the problem of having insufficientchannel data for analysis.\\

\begin{figure}[t]
	\centering
	\includegraphics[width=0.4\textwidth]{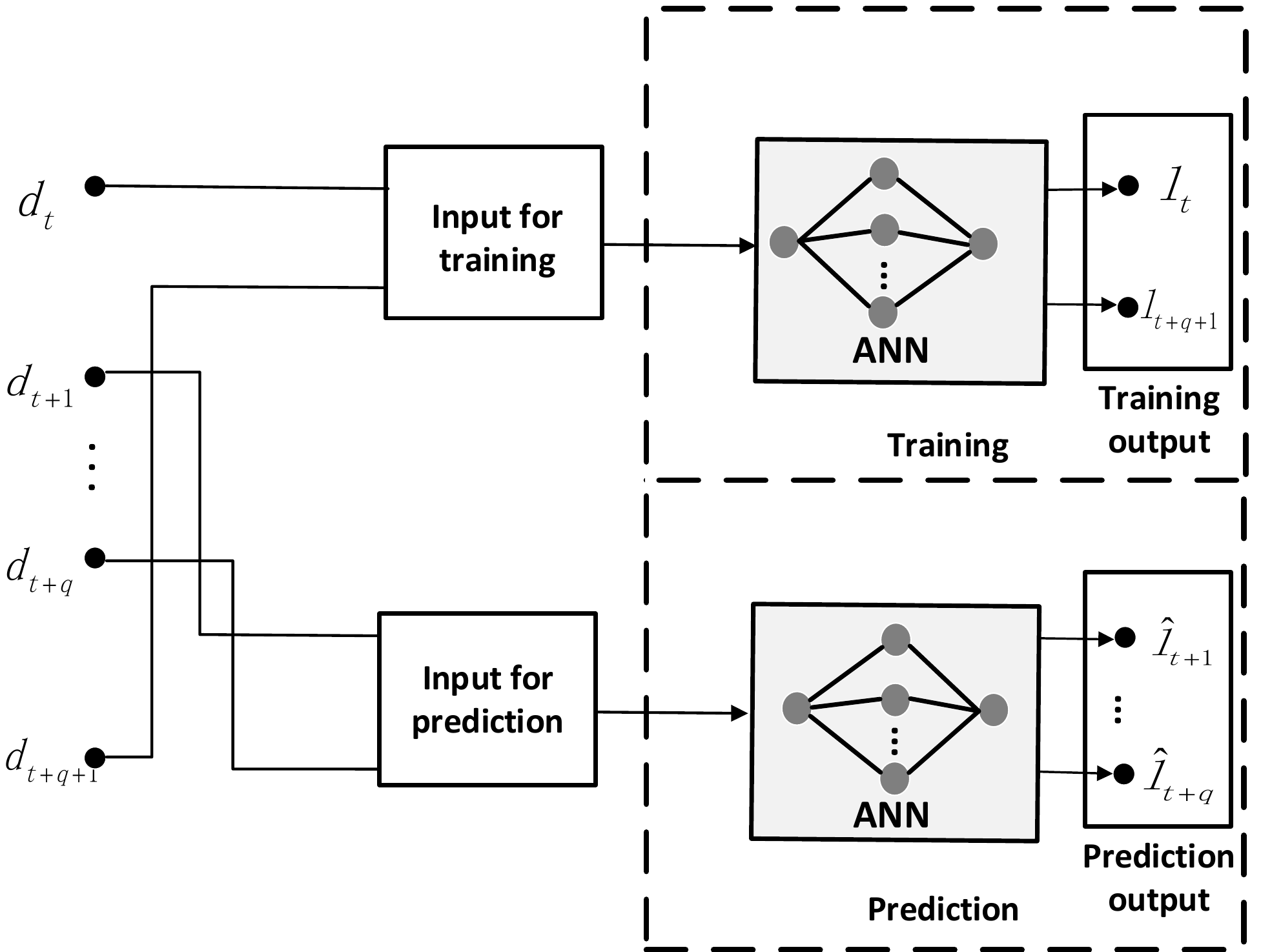}
	\caption{Process of training and prediction}
	\label{fig1}
\end{figure}

This paper proposes a channel prediction algorithm, which trains ANN to learn channel history and predict path loss and shadow fading at different distances. Fig. \ref{fig1} illustrates the framework of the proposed algorithm. In the training process, $ d_{t} $ and $ d_{t+q+1} $ are the distances of the measured channel data in time $\mathit{t}$ and $\mathit{t+q}+1$, respectively, whereas $ \mathit{q} $ is the number of the predicted points. Parameters $ l_{t} $ and $ l_{t+q+1} $ are network output in training process (including path loss and shadowing). After the training, the distances of those unmeasured points (from $ d_{t+1} $ to $ d_{t+q} $) are used as network input, and the predicted channel data (from $ \hat{l}_{t+1} $ to $ \hat{l}_{t+q} $) are obtained as the output of ANN. It is found that the proposed approach can achieve fairly high accuracy in channel prediction, which can be used to improve channel characterization when measurements are insufficient.\\

\section{Channel Measurement}

For ANN training, measurement data is needed. In this work, channel measurements were carried out in railway mobile scenario. The measurements were carried out at 460 MHz with a bandwidth of 20 MHz. The sample interval $ T_{r} $ is 5.12 ms which satisfies the Nyquist criterion for mobile channel measurements\cite{M2011}. In measurements, the inteval of distance between each data point is 1.42 m. The total number of measurement data is 3000. More details of the measurements can be found in \cite{wen}.\\

The measurement system records channel transfer functions $ \mathit{H(d,f_{l})} $, where $ \mathit{d} $ is the distance between Tx and Rx antennas and $ f_{l} $ is the $ \mathit{l} $ th frequency point. The power of the received signal $ P_{r}(d) $ is determined from $ \mathit{H(d,f_{l})} $ as

\begin{equation}\label{PG}
	P_{r}(d)=\dfrac{1}{N_{f}}(\sum_{l=1}^{N_{f}}\left|H(d,f_{l})\right|  ^{2})
\end{equation}
where $ \mathit{N_{f}} $ is the total number of the measured frequency points, and in the measurements, $ \mathit{N_{f}}=1024 $.\\

The raw channel large-scale component $ \mathit{PL} $, including both path loss and large-scale fading (LSF), can be calculated as follows

\begin{equation}\label{PL}
	PL(dB)=P_{t}+G_{T_{X}}+G_{R_{X}}-P_{r}
\end{equation}

To remove the components of small-scale fading, $ \mathit{PL} $ is averaged with a sliding window of 40 wavelength. To extract LSF component $ \mathit{X_{\sigma}}(dB) $, the linear model of path loss should be calculated first, and then the LSF $  \mathit{X_{\sigma}}$ can be obtained as\\
\begin{equation}\label{LSF}
	\mathit{X_{\sigma}}(dB)=PL(dB)-\bar{PL}(dB)
\end{equation}
where $ \overline{PL} $ is the linear model of path loss.

\section{Channel Prediction with ANN}

\begin{figure}[t]
	\centering
	\includegraphics[width=0.5\textwidth]{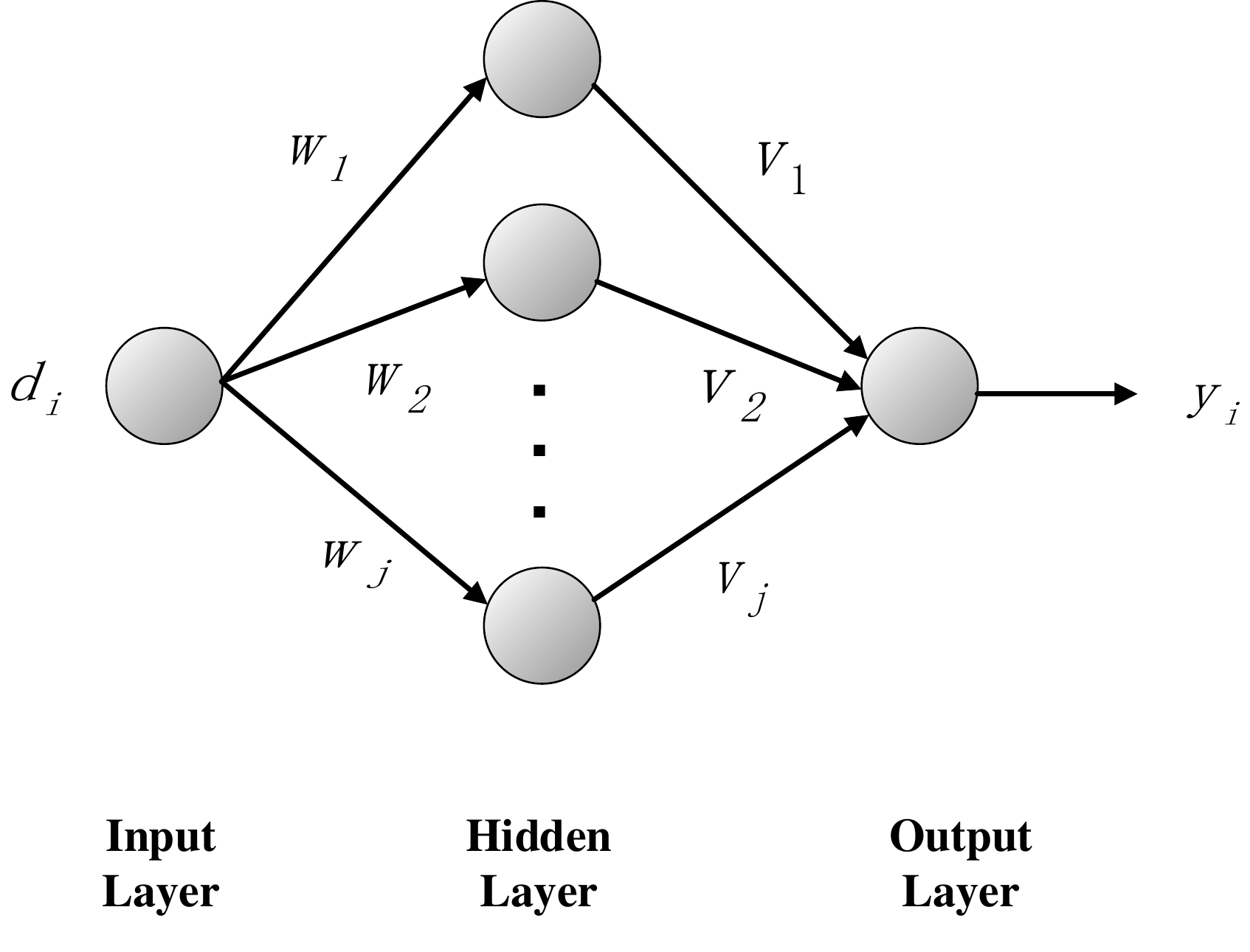}
	\caption{Structure of ANN}
	\label{fig2}
\end{figure}

The ANN includes many types of neural networks such as back propagation network (BPN), RBF-NN, extreme learning machine (ELM), etc. They have been widely used for different applications, including pattern recongnition, image processing, forcasting, automatic control, etc\cite{plrailway}. In this paper, the BPN, RBF-NN and ELM are chosen due to the good performances in interpolation prediction. Those methods can well realize complex nonlinear mapping and have good self-learning ability. The basic architecture of the three kinds of ANN are shown in Fig. \ref{fig2}. The number of hidden layer and the number of neurons in input layer and output layer are all set to be 1 in this paper. The output of the ANN can be expressed as

\begin{equation}\label{BPN}
	y_{i}=F_{o}(\sum_{j=1}^{M}v_{j}F_{n}(w_{j}d_{i}))
\end{equation}
where $ d_{i} $ is the $\mathit{i}$ th input data, and in this paper it represents distance $ \mathit{d} $. Parameter $ y_{i} $ is the $\mathit{i}$ th output data, and in this paper it represents the predicted $ \mathit{PL} $. Parameter $ F_{o}(\cdot) $ is the activation function of the output layer, and $ F_{n}(\cdot) $ is the activation function of the hidden layer, Parameter $ v_{j} $ represents the synaptic weight from the $ \mathit{j} $ th neuron in the hidden layer to the output layer, $ w_{j} $ represents the synaptic weight from the input layer to the $ \mathit{j} $ th neuron in the hidden layer, and $ \mathit{M} $ is the number of the hidden neurons. For BPN and ELM, $ v_{j} $ and $ w_{j} $ are decided by training process. For RBF-NN, the value of $ w_{j} $ always equals to $ 1 $, and $ v_{j} $ is decided by training process. For channel prediction, some measurement data are equally-spaced chosen as training data set to predict the rest data.  After the ANN has been trained successfully, the distances that needs to be predicted are input to the ANN to obtain channel data.\\

In the training process of BPN, the core is the back propagation algorithm to adjust network weights to reduce the deviation between network output and measurement data, which can be expressed as
\begin{equation}\label{ek}
	E_{i}=\frac{1}{2}\sum(y_{i^{'}}-y_{i})^{2}
\end{equation}

\begin{equation}\label{d1}
	\Delta v_{j}=-\eta\frac{ \partial E_{i}}{ \partial v_{j}}
\end{equation}



\begin{equation}\label{d4}
	\Delta v_{j}=(y_{i}-y_{i^{'}})y_{i}(1-y_{i})b_{j}
\end{equation}
where $y_{i^{'}}$ is the $\mathit{i}$ th measurement data and $E_{i}$ is the current predictive error. Parameter $ \eta =10^{-6} $ is the learning rate of BPN and $\Delta v_{j}$ is adjusted value of weight base on the current $E_{i}$. Parameter $b_{j}$ is output of the $ \mathit{j} $ th neuron. With the back propagation algorithm, using the current predicted result to adjust synaptic weights is possible. Each round of prediction and back propagation is called an iteration. When $ E_{i} $ is smaller than the error threshold ($ 10^{-5} $) or the number of iterations reaches the preset upper limit, the training process is finished, and the maximum number of iterations is set to be 1000 in order to prevent the error always being larger than the error threshold.\\

The ELM abandons the iteration process of BPN training. When $ w_{j} $ is set randomly, the function of $ w_{j} $ and $ v_{j} $ can be determined by linear equations.\\

The RBF-NN focuses on extracting partial characteristics of data. The RBF-NN finds $ \mathit{M} $ central data points by K-Means. From each central point, the activation function can be determined by a Gaussian function and the weights between hidden layer and output layer are also determined by linear equations.

\begin{figure*}[t]
	\centering
	\subfloat[]{\includegraphics[width=0.4\textwidth]{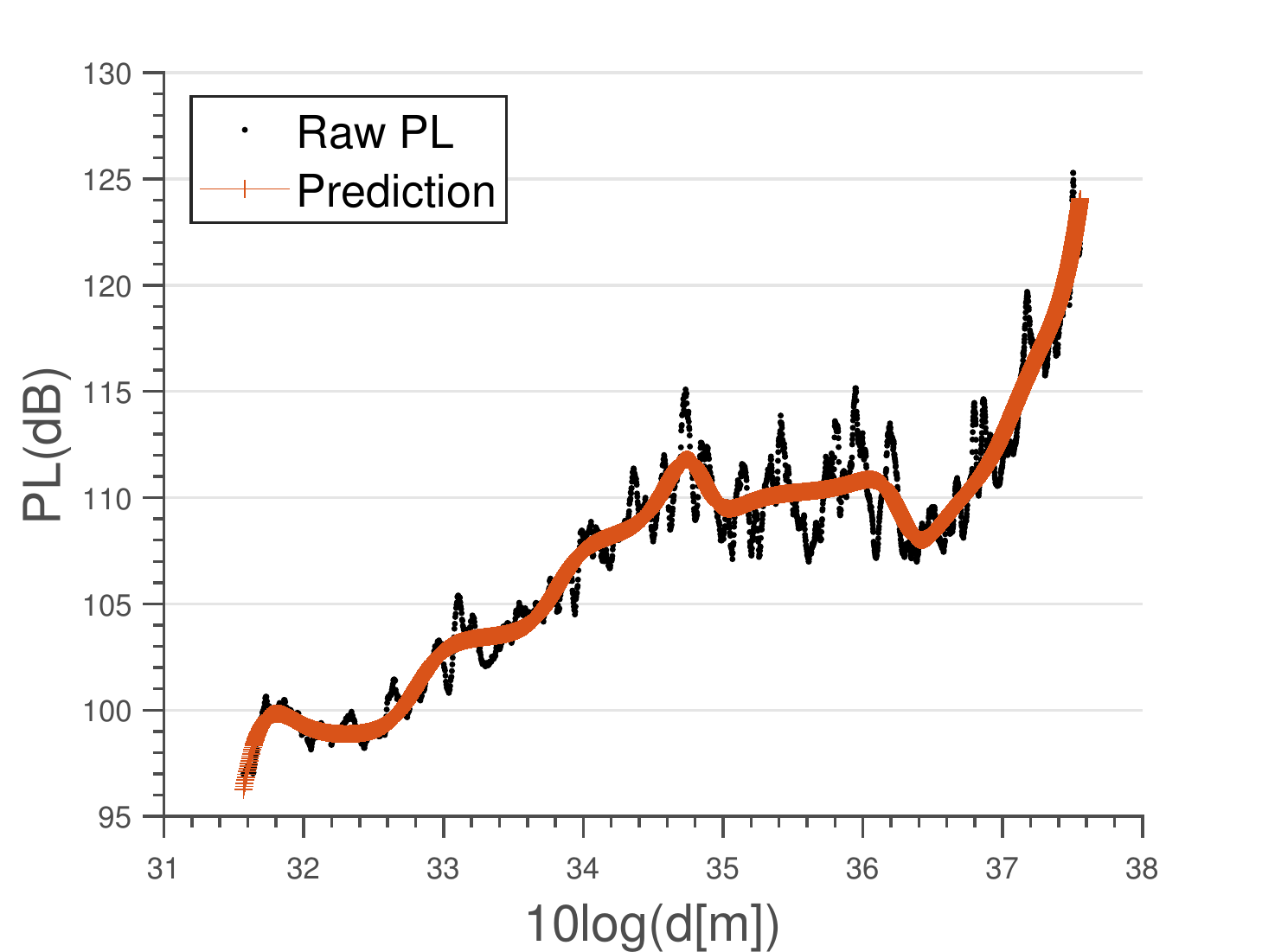}%
		\label{fig3a}}
	\subfloat[]{\includegraphics[width=0.4\textwidth]{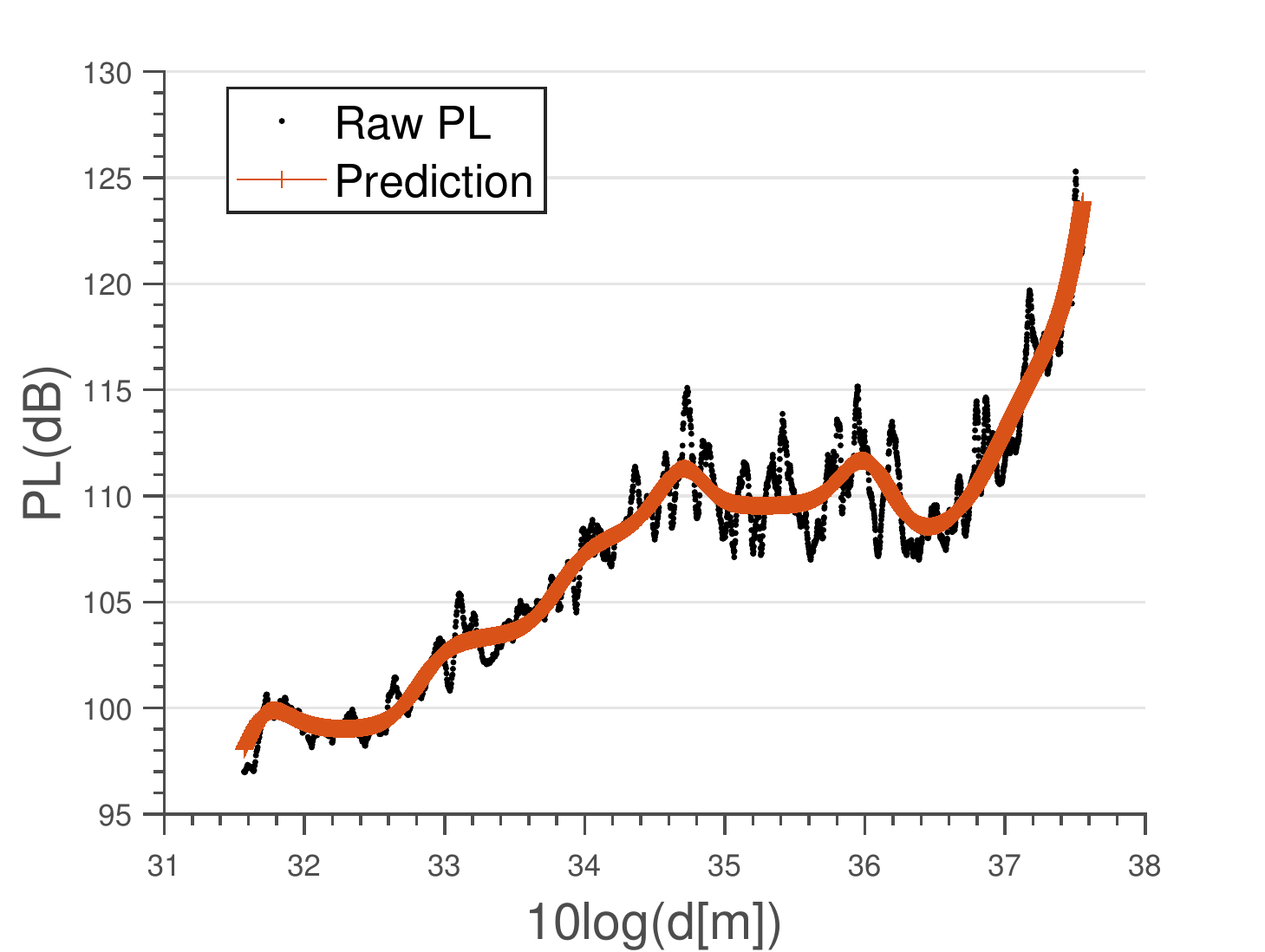}%
		\label{fig3b}}
	\hfil
    \subfloat[]{\includegraphics[width=0.4\textwidth]{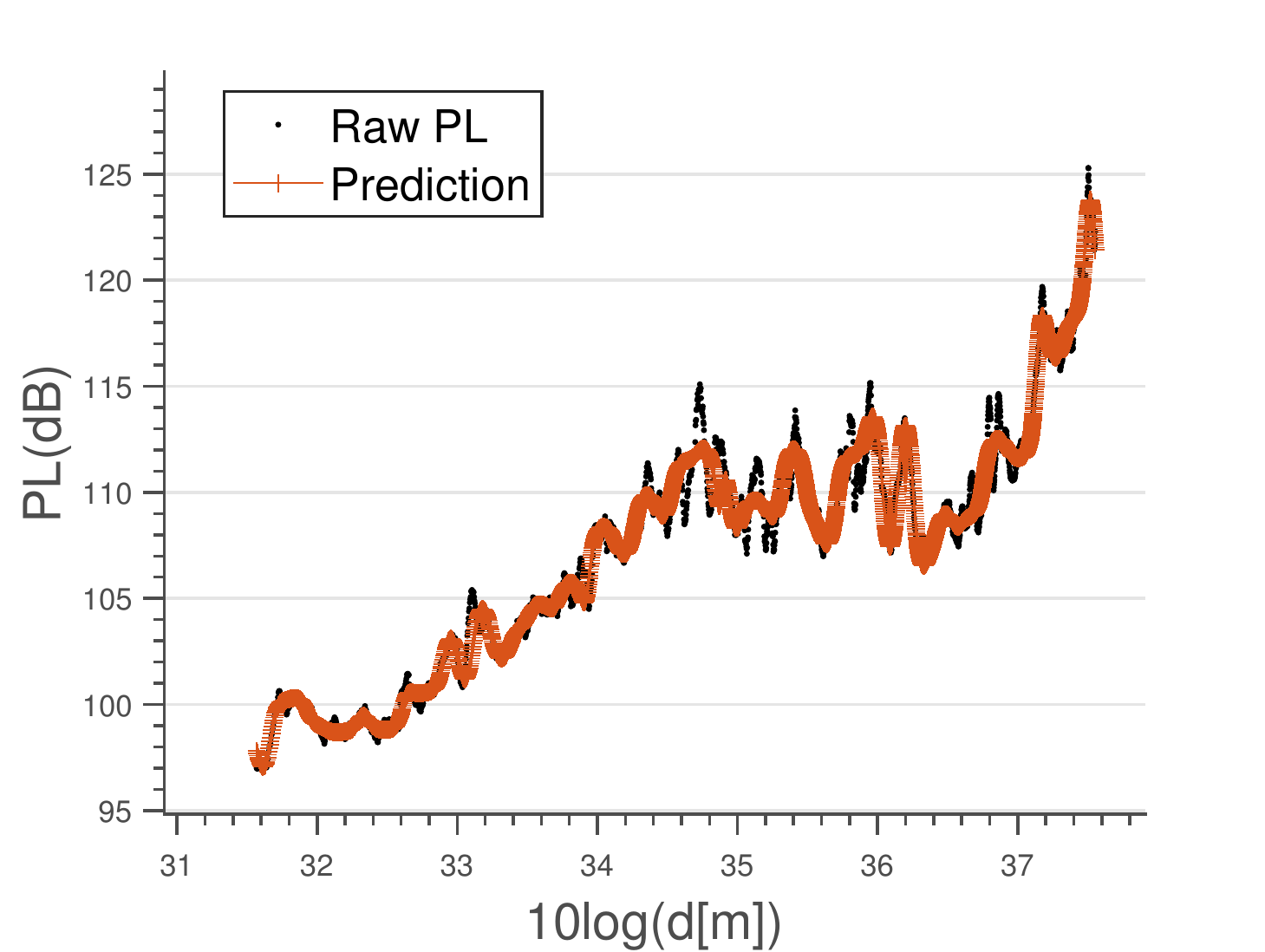}%
		\label{fig3c}}
	\subfloat[]{\includegraphics[width=0.4\textwidth]{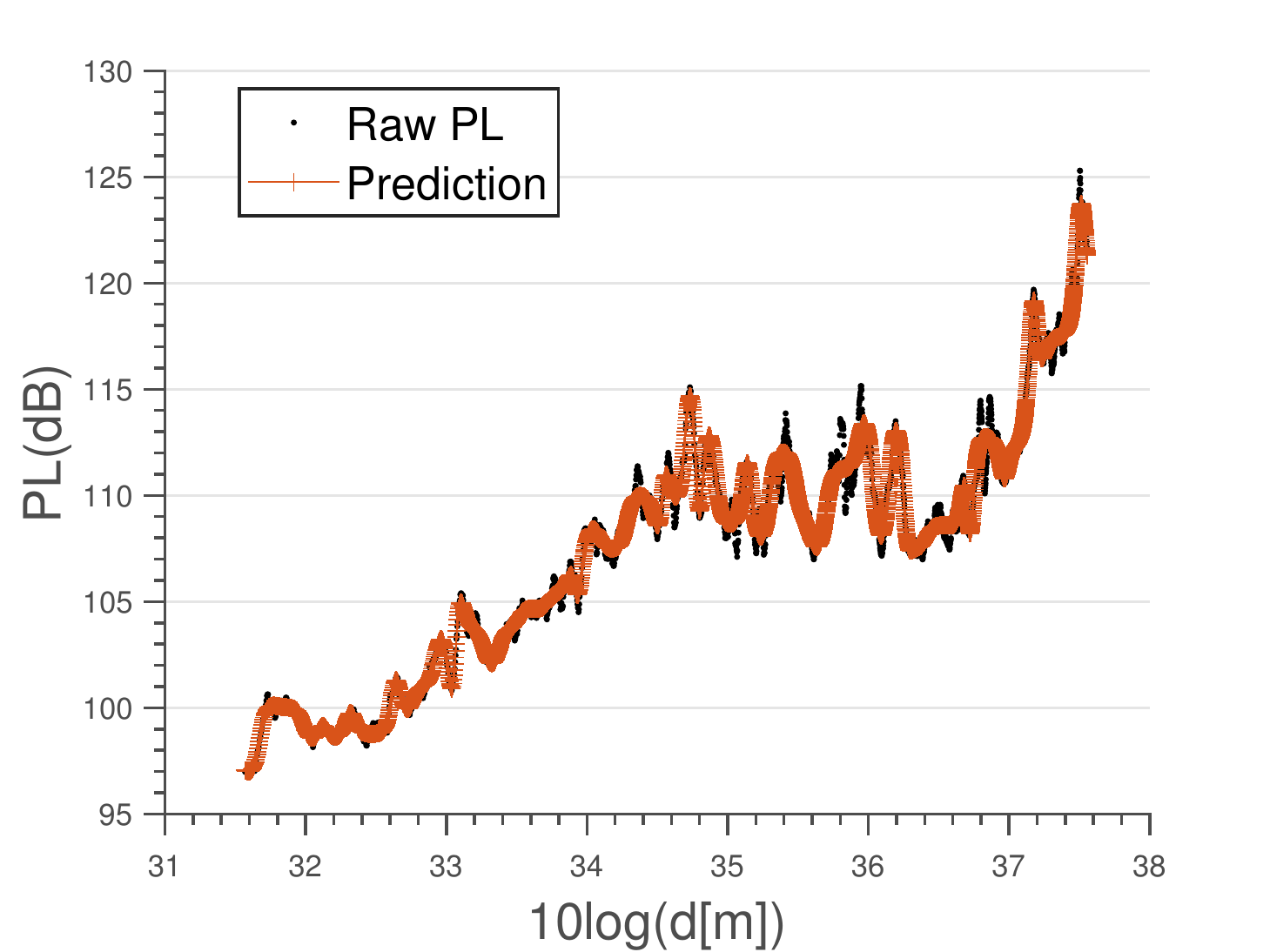}%
		\label{fig3d}}
	\caption{Prediction results of $ \mathit{PL} $ by BPN. (a) 14$\%$ training ratio and 10 neurons; (b) 50$\%$ training ratio and 10 neurons; (c) 14$\%$ training ratio and 50 neurons; (d) 50$\%$ training ratio and 50 neurons}
	\label{fig3}
\end{figure*}
\section{Simulation Result}

In order to verify the proposed prediction method and explore the possible factors that could affect prediction performance, some simulatons are conducted. In the simulations, different proportions of training data set and number of neurons are chosen for analysis. For the proportions of training data set, it can be expressed as $ \mathit{r=\frac{1}{q+1}}  $ where $ \mathit{q} $ is the number of predicted data between the adjacent measurement data as shown in Fig. \ref{fig1}. It can be seen that $ \mathit{r} $ will not be more than 50$ \% $. For the number of neurons, the conditions of 10, 20, 30, 40 and 50 neurons are considered in simulations. In this paper, the single layer ANN is considered, which is found to have fairly high accuracy and low complexity.\\

Fig. \ref{fig3} shows the simulations of having 14$ \% $ (with 6 prediction data between each adjacent training data) and 50 $ \% $ (with 1 prediction data between each adjacent training data) training ratios with the BPN that constituted by 10 and 50 neurons, respectively. The black data points are the measurement data; the red data points are the output data for comparison, which include the training data set and predicted data. In Fig. \ref{fig3}, it is found that the predicted results are generally in good agreement with the measurements. It is also found that the number of neurons has larger impact on prediction result than the proportion of training data set. A larger number of neurons can significantly improve prediction, whereas the improvement of using higher proportion of training data set is relatively small.\\

\begin{figure}[t]
	\centering
	\includegraphics[width=0.45\textwidth]{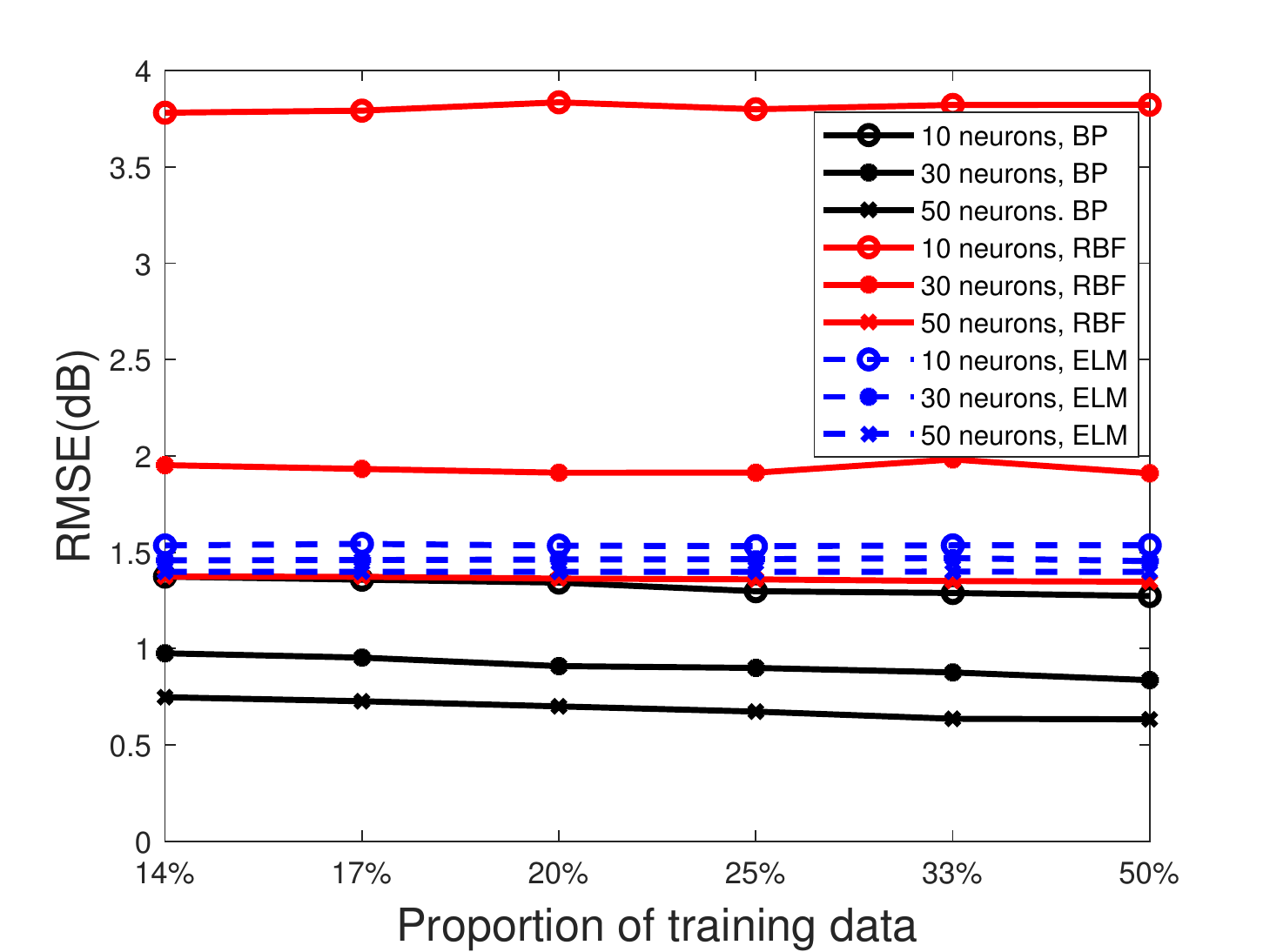}
	\caption{RMSE of $ \mathit{PL} $ prediction}
	\label{fig4}
\end{figure}

\begin{figure*}[t]
	\centering
	\subfloat[]{\includegraphics[width=0.4\textwidth]{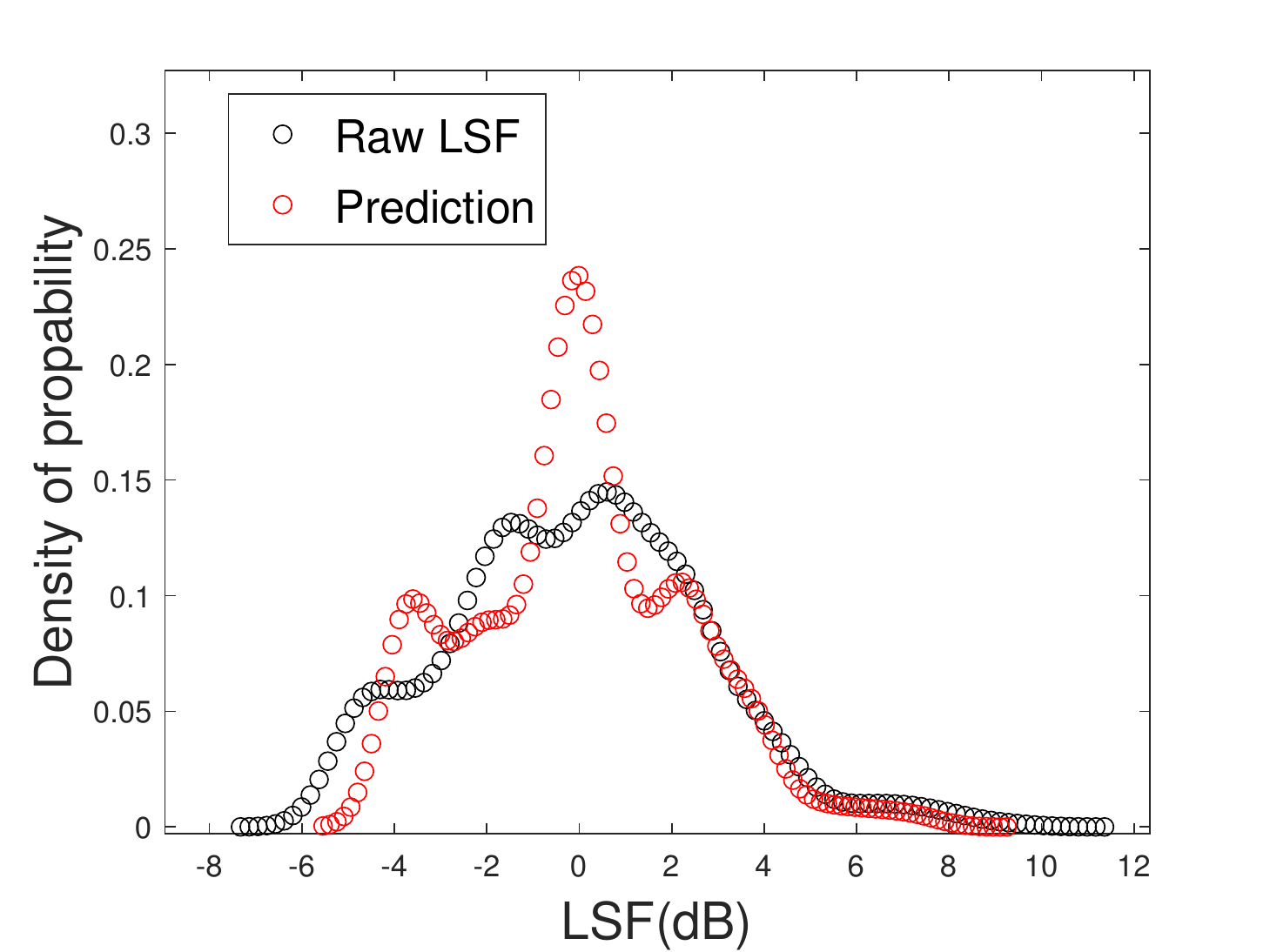}%
		\label{fig6a}}
	\subfloat[]{\includegraphics[width=0.4\textwidth]{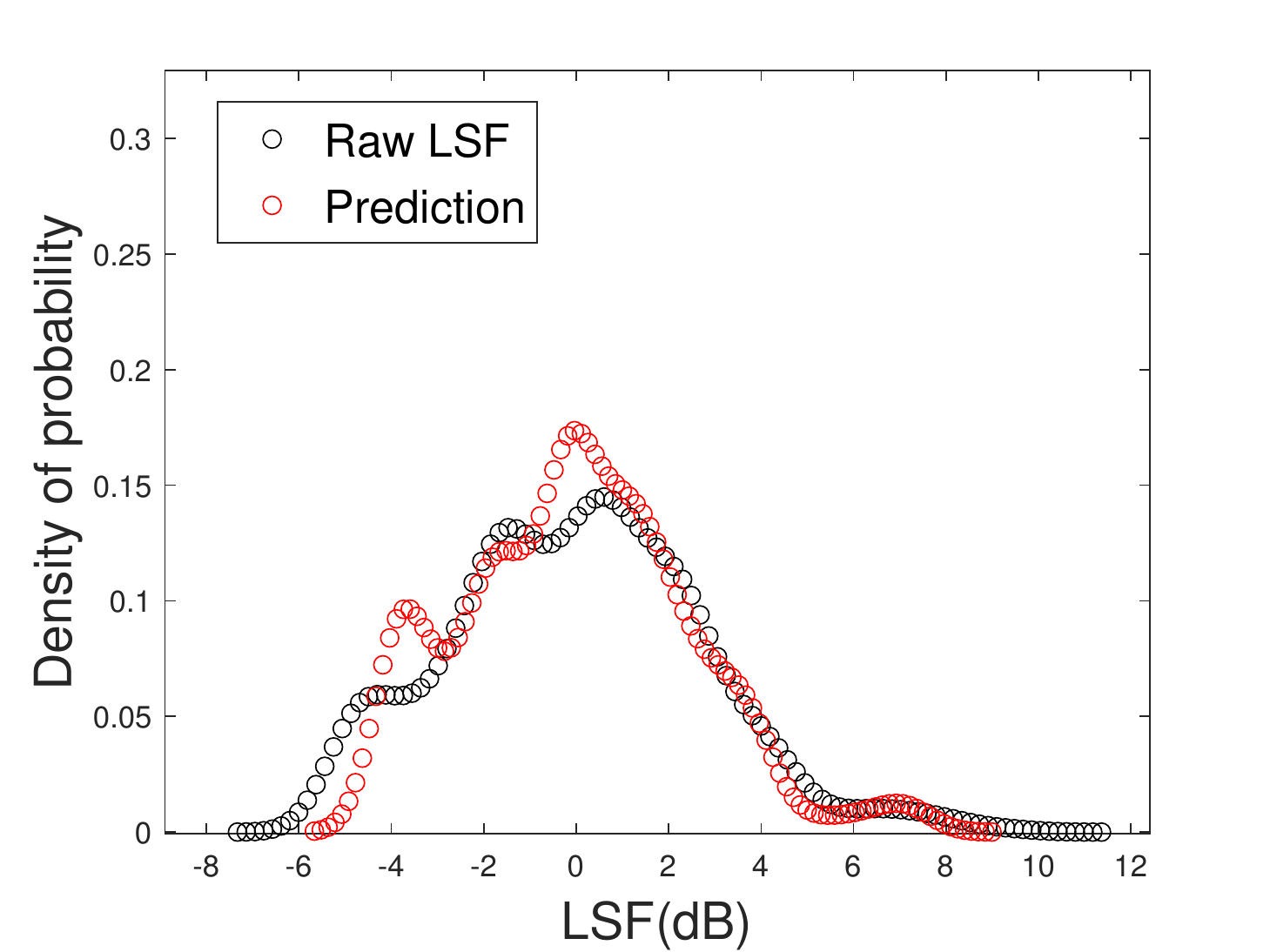}%
		\label{fig6b}}
	\hfil
	\subfloat[]{\includegraphics[width=0.4\textwidth]{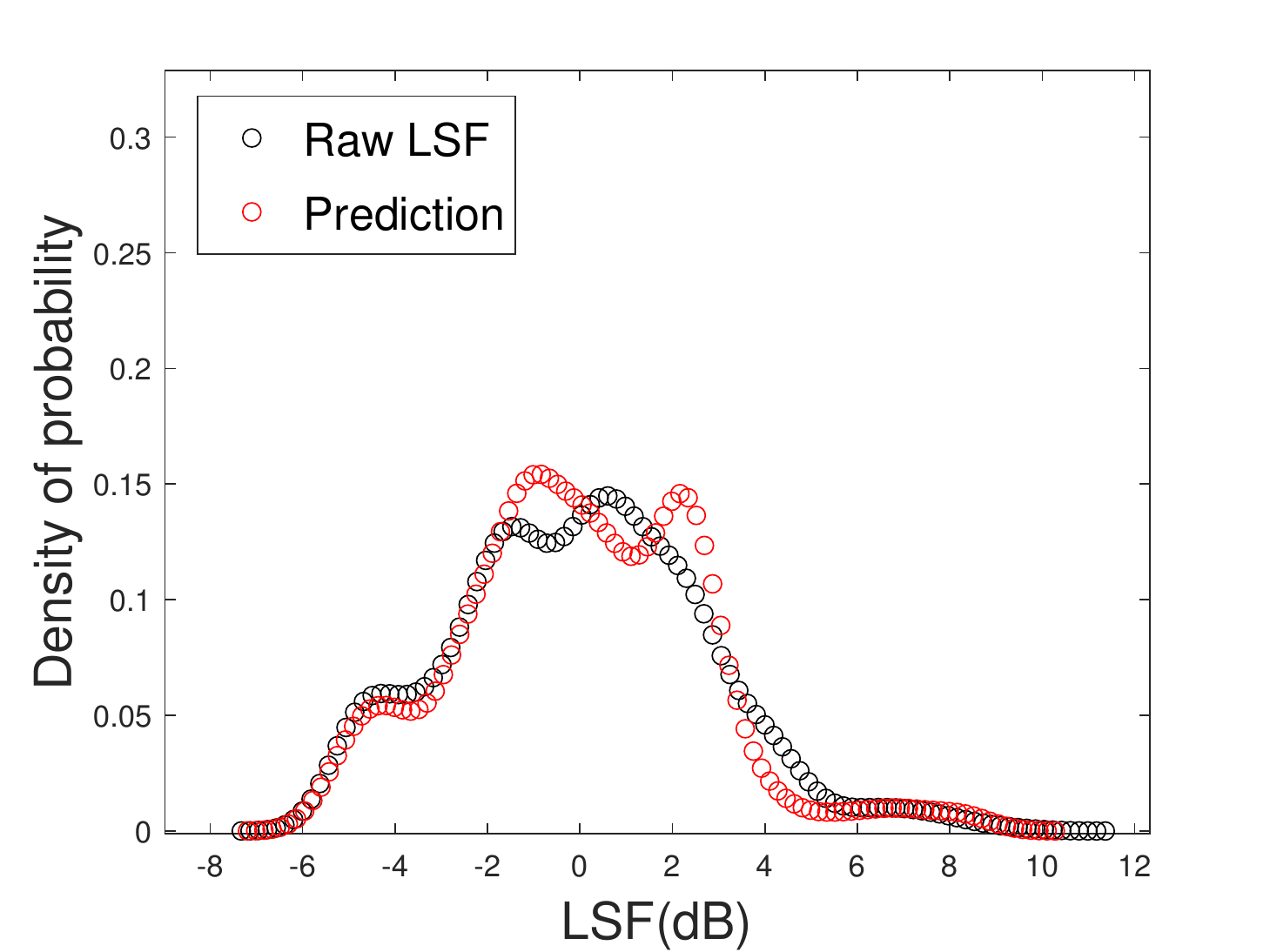}%
		\label{fig6c}}
	\subfloat[]{\includegraphics[width=0.4\textwidth]{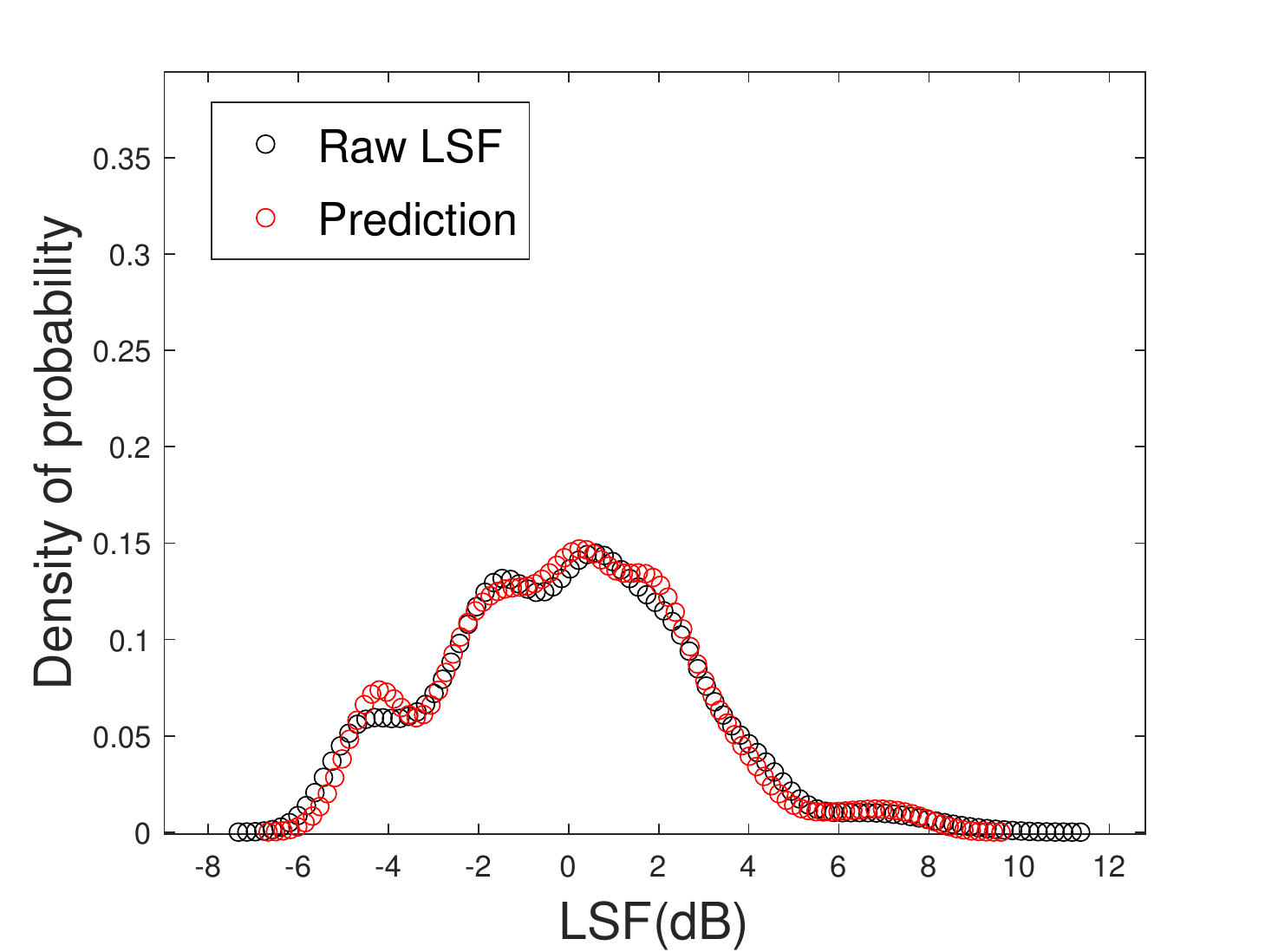}%
		\label{fig6d}}
	\caption{Predicting result of LSF by BPN. (a) 14$\%$ training ratio and 10 neurons; (b) 50$\%$ training ratio and 10 neurons; (c) 14$\%$ training ratio and 50 neurons; (d) 50$\%$ training ratio and 50 neurons}
	\label{fig6}
\end{figure*}
To better evaluate prediction accuracy, the Root-Mean-Square error (RMSE) $ \mathit{R} $ is calcaulated as

\begin{equation}\label{RMSE}
	\mathit{R}=\sqrt{\dfrac{\sum_{i=1}^{N}(m_{i}-p_{i})^{2}}{N}}
\end{equation}
where $ m_{i} $ is $ \mathit{i} $ th $ \mathit{PL} $ from measurements, $ p_{i} $ is the predicted result of $ \mathit{PL} $, and $ \mathit{N} $ is the total number of measurement data points.\\

In Fig. \ref{fig4}, the RMSE of predicted results using different types of networks are shown. It is found that the BPN has the best performence, and the performances of ELM and RBF-NN are generally worse. The mainly advantage of ELM is running speed, which uses a simplified training process to reduce calcaulation time. The running speed of RBF-NN is between ELM and BPN, because RBF-NN adopts partial learning. It is also observed in Fig. \ref{fig4} that for a specific network, a high proportion of training data set leads to low prediction error. However, the impact of the proportion of training data is fairly small. For most cases, RMSE reduces less than 0.2 dB when the proportion increases from 14$ \% $ to 50$ \% $, as shown in Fig. \ref{fig4}. Compared with the impact of proportion of training data set, it is found that the decreasing of RMSE caused by more neurons is much more distinct, especially for BPN and RBF. It can be seen from Fig. \ref{fig4} that the RMSE of 50 neurons case is generally 1 dB less than the case of 10 neurons in BPN. Therefore, when ANN is used to predict $ \mathit{PL} $, more neurons significantly improve prediction accuracy. Moreover, Fig. \ref{fig4} shows that ANN can be well used for channel prediction. Even for the case with few training data (i.e., the proportion of training data is low), the accuracy is fairly high compared with using a large number of training data (e.g., the proportion of training data is 50$ \% $).\\

\begin{figure}[t]
	\centering
	\includegraphics[width=0.45\textwidth]{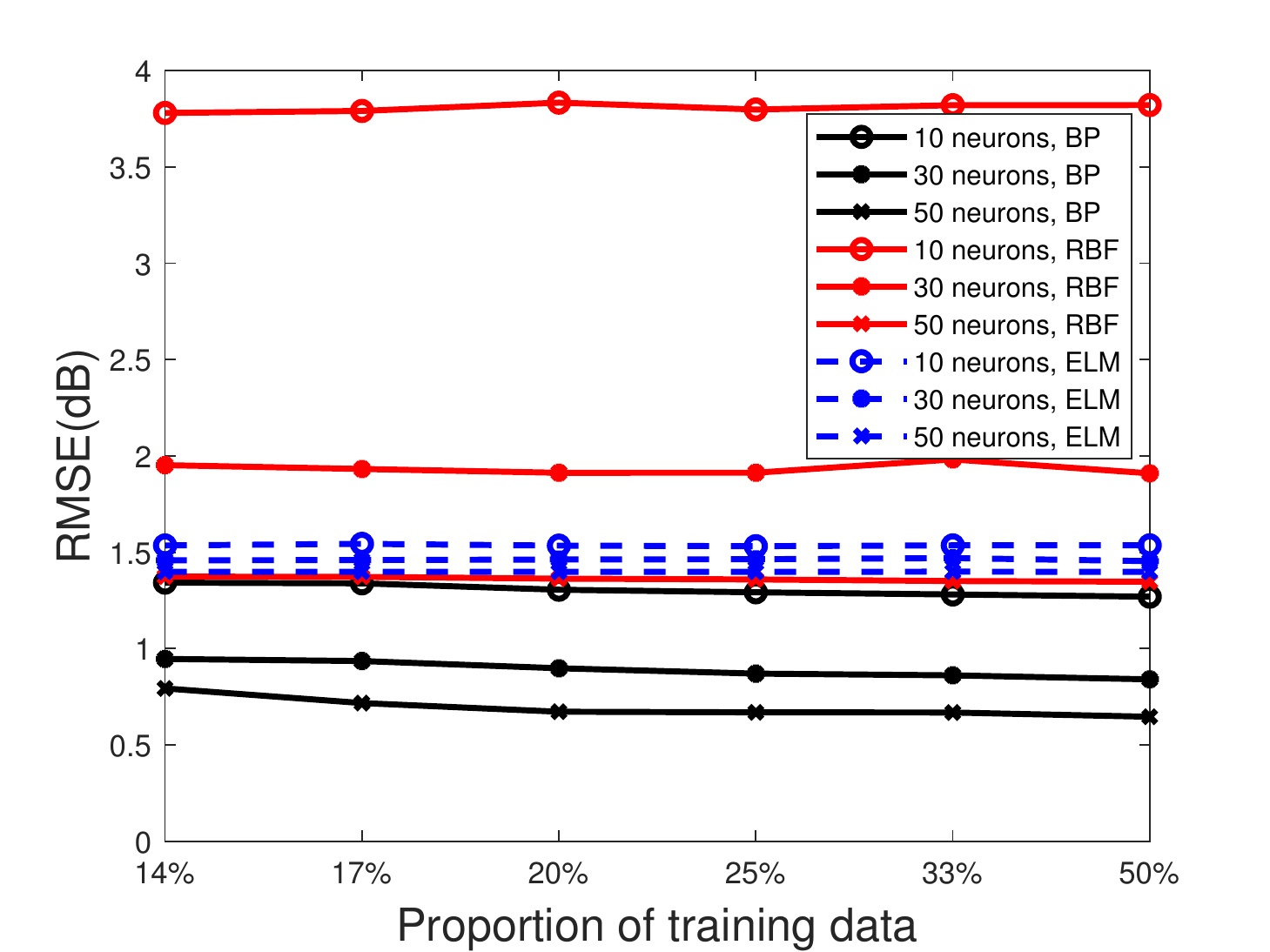}
	\caption{RMSE of LSF prediction}
	\label{fig5}
\end{figure}

For LSF prediction analysis, the zero mean Gaussian distribution is chosen as the model of LSF, and the density of probability is used for comparison between the prediction and measurements. The results of BPN are shown in Fig. \ref{fig6}, where the black circles represent the probability density from measurement data and the red circles represent the probability density from predicted data. It is observed that more neurons and higher $ \mathit{r} $ can significantly improve the statistical distribution fitting of LSF, especially in Fig. \ref{fig6}(d). Fig. \ref{fig5} shows the RMSE results of LSF prediction. It is found that the trend of curves are generally similar with the results in Fig. \ref{fig4}. Moreover, the values of RMSEs of the predicted $ \mathit{PL} $ and LSF are also fairly close to each other. It is concluded that the BPN has the best performance for LSF prediction, and using more neurons significantly improves prediction accuracy.

\section{Conclusion}

In this paper, a scheme of using ANN to predict channel path loss and large-scale fading is proposed. The ANN can extract the characteristic of channel and predict the channel data of the neighboring unmeasured points. The performances of using different types of ANN are compared. The BPN is found to have the smallest prediction error, the error of ELM is slightly larger than BPN and the RBF-NN is found to have the highest channel prediction error.  Moreover, it is found that the number of neurons and the proportion of training data can effect the error of prediction. Using more number of neurons can significantly reduce prediction error and the influence of proportion of training data is relatively small.

%
%


\begin{thebibliography}{9}

\bibitem{C} J. Zhang, P. Tang, L. Yu, T. Jiang, and L. Tian, “Channel measurements and models for 6G: current status and future outlook,” \emph{Frontiers of Information Technology Electronic Engineering}, 21, 2020, pp. 39–61.

\bibitem{Lee} W. C. Y. Lee, “Estimate of local average power of a mobile radio signal,” \emph{IEEE Transactions on Vehicular Technology}, \textbf{34}, 1, Febuary 1985, pp. 22–27.

\bibitem{w} J. Huang, C. Wang, L. Bai, J. Sun, Y. Yang, J. Li, O. Tirkkonen, and M. Zhou, “A big data enabled channel model for 5G wireless
communication systems,” \emph{IEEE Transactions on Big Data}, \textbf{6}, 2, 2020, pp. 211–222.

\bibitem{w44} Po-Rong Chang and Wen-Hao Yang, “Environment-adaptation mobile radio propagation prediction using radial basis function neural networks,” \emph{IEEE Transactions on Vehicular Technology}, \textbf{46}, 1, 1997, pp. 155–160.

\bibitem{w47} L. Azpilicueta, M. Rawat, K. Rawat, F. M. Ghannouchi, and F. Falcone, “A Ray Launching-Neural Network approach for radio wave propagation analysis in complex indoor environments,” \emph{IEEE Transactions on Antennas and Propagation}, \textbf{62}, 5, 2014, pp. 2777–2786.

\bibitem{w48} J. Zhang, “The interdisciplinary research of big data and wireless channel: A cluster-nuclei based channel model,” \emph{China Communications}, \textbf{13}, 2, 2016, pp. 14–26.

\bibitem{w51} G. P. Ferreira, L. J. Matos, and J. M. M. Silva, “Improvement of outdoor signal strength prediction in UHF Band by Artificial Neural Network,” \emph{IEEE Transactions on Antennas and Propagation}, \textbf{64}, 12, 2016, pp. 5404–5410.

\bibitem{w53} T. Blazek and C. F. Mecklenbr¨auker, “Sparse time-variant impulse response estimation for vehicular channels using the c-LASSO,” 2017 IEEE 28th Annual International Symposium on Personal, Indoor, and
Mobile Radio Communications (PIMRC), 2017, pp. 1–5.

\bibitem{w54} M. Uccellari, F. Facchini, M. Sola, E. Sirignano, G. M. Vitetta, A. Barbieri, and S. Tondelli, “On the application of support vector machines to the prediction of propagation losses at 169 MHz for smart metering applications,” \emph{IET Microwave Antennas Propagation}, \textbf{12}, 3, 2018, pp. 302–312.

\bibitem{w45} Z. Liu, Z. Huang, and Y. Zhou, “An efficient maximum likelihood method for Direction-of-Arrival estimation via Sparse Bayesian Learning,” \emph{IEEE Transactions on Wireless Communications}, \textbf{11}, 10, 2012, pp. 1–11.

\bibitem{w55} R. He, B. Ai, A. F. Molisch, G. L. Stuber, Q. Li, Z. Zhong, and J. Yu, “Clustering enabled wireless channel modeling using big data
algorithms,” \emph{IEEE Communications Magazine}, \textbf{56}, 5, 2018, pp. 177–183.

\bibitem{M2011} A. F. Molisch, \emph{Wireless Communications}, New York, IEEE-Wiley, 2011.

\bibitem{wen} Z. R. Wen, R. S. He, B. Ai, B. Zhang, M. Yang, W. Wang, Z. D. Zhong, and H. X. Zhang, “Measurement and modeling of LTE-railway channels in high-speed railway environment,” \emph{Radio Science}, \textbf{55}, 4, 2020, pp. 1–12.

\bibitem{plrailway} D. Wu, G. Zhu, and B. Ai, “Application of Artificial Neural Networks for path loss predictio in railway environments,” 2010 5th International ICST Conference on Communications and Networking in China, 2010, pp. 1–5.


\end{thebibliography}
\end{document}